\documentclass[prx,aps,superscriptaddress,nofootinbib,twocolumn]{revtex4-1}
\usepackage{mathrsfs}
\usepackage{amsfonts}
\usepackage{amssymb}
\usepackage{amsmath}
\usepackage{graphicx}
\usepackage[usenames,dvipsnames]{color}
\usepackage[colorlinks=true,citecolor=blue,linkcolor=magenta]{hyperref}
\usepackage{ulem}
\usepackage{lmodern}
\usepackage{enumitem}
\usepackage{multirow}
\usepackage{soul}
\usepackage{bbm}
\usepackage{algorithm}
\usepackage{algpseudocode}

\begin{document}

\title{A high threshold code for modular hardware with asymmetric noise}

\author{Xiaosi Xu}
\affiliation{Department of Materials, University of Oxford, Parks Road, Oxford OX1 3PH, United Kingdom}

\author{Qi Zhao}
\affiliation{Center for Quantum Information, Institute for Interdisciplinary Information Sciences, Tsinghua University, Beijing 100084, P.~R.~China}

\author{Xiao Yuan}
\author{Simon C. Benjamin}
\affiliation{Department of Materials, University of Oxford, Parks Road, Oxford OX1 3PH, United Kingdom}

\begin{abstract}

{We consider an approach to fault tolerant quantum computing based on a simple error detecting code operating as the substrate for a conventional surface code. 
We develop a customised decoder to process the information about the likely location of errors, obtained from the error detect stage, with an advanced variant of the minimum weight perfect matching algorithm.
A threshold gate-level error rate of 1.42\% is found for the concatenated code given highly asymmetric noise. This is superior to the standard surface code and remains so as we introduce a significant component of depolarising noise; specifically, until the latter is 70\% the strength of the former. Moreover, given the asymmetric noise case, the threshold rises to 6.24\% if we additionally assume that local operations have 20 times higher fidelity than long range gates. Thus for systems that are both modular and prone to asymmetric noise our code structure can be very advantageous. }
\end{abstract}

\maketitle

\section{Introduction}
\label{Introduction}
To realise the promise of large-scale quantum computers that outperform classical machines, a protective mechanism must be employed as quantum states are fragile and easily damaged by their noisy environment. 
Many error correction codes have therefore been developed, that can detect and correct errors. In essence, a group of physical qubits (such as ions, superconducting elements, etc) are used to collectively store a smaller number of `logical' qubits.   The group of topological codes~\cite{dennis2002topological, raussendorf2007topological} are particularly attractive solutions because they typically have a local structure for the {\it stabiliser} check operations that must be performed to identify errors. This leads to relatively simple protocols supporting fault-tolerant quantum computing, i.e, ensuring a single error occurring during an error-correction cycle will not itself corrupt the logical qubit(s) that are the subject of the cycle. This leads to a {\it threshold} rate for errors at the physical level~\cite{aharonov2008fault, raussendorf2007fault}; when error rates are within this threshold, one can achieve an arbitrarily low {\it logical error rate} by having a suitable large ratio of physical qubits to logical qubits. The surface code, as a 2D topological code, is regarded as a highly promising code due to its modest requirement of a 2D nearest-neighbour connectivity, a high threshold (resulting from low degree stabiliser checks), and simple grid-like lattice structure~\cite{kitaev1997quantum,kitaev2003fault,fowler2012surface}. As the surface code deals with bit-flip and phase-flip errors independently, we can perform checks for $X$-type and $Z$-type errors alternatively. However, in practice many systems experience an asymmetric error source that makes the standard surface code no longer the optimal choice; relevant cases are reported in Refs.~\cite{aliferis2009fault, lanyon2011universal, saeedi2013room}. 

In this paper we consider the scenario where phase errors are more prevalent than bit-flip errors (it immediately applies to the converse case where bit-flip, rather than phase, is prevalent). Commonly in real systems the noise processes are complex, involving both environmental elements and aspects due to the active gates, but generally phase processes take place with a different severity to flip processes~\cite{schmidt2003realization,zeiher2015microscopic,yoneda2018quantum}. Note such a case has been considered in recent theoretical studies~\cite{tuckett2018ultrahigh,li20182}, where high thresholds were found for extremely asymmetric noise from the environment while the active gate-level operations of the computer were presumed perfect. Here we consider a range of noise models (degrees of asymmetry) and moreover we track noise events up from gate-level events, all of which are assumed imperfect to some degree. Our approach is to introduce a variant of the canonical surface code by concatenating it with a two-qubit phase detection code. Thus the `data qubits' of the surface code are no longer physical qubits, but rather are qubit pairs in the phase-detecting code (this can be trivially adjusted to the bit-flip-detecting code if instead bit-flip errors are prevalent). We note that a similar case of the surface code concatenated with the full $[[4,4,2]]$ error detecting code has been considered by \textcite{criger2016noise}, who note the equivalence to variants of other topological codes and thus obtain an estimate of the threshold (however, for the case of gate-level errors, this is lower than the canonical surface code threshold). Another comparable work has considered a cluster-state topological code concatenated with the repetition code and observed a marked threshold gain in the case of extremely biased gate-level noise~\cite{stephens2013high}. Here our base-level code is simply the 2D surface code and moreover, rather than mapping to an equivalent code, we 
propose a new surface code decoder (the algorithm that attempts to infer optimal error-correcting operations) by feeding in the information on likely error locations obtained by the lower-level error detection. We find this results in significantly enhanced thresholds, as presently noted.

The error-correction cycle in the present study is composed of two phases: (1) Local checks where we measure the $XX$ stabiliser for the pairs of physical qubits, each such pair constituting an individual surface code data qubit, and (2) Parity checks for each unit of four data qubits, as per the normal surface code, but here of course this must be performed in such as fashion as to respect the lower level code. By concatenation we double the number of physical qubits, and thus increase the number of gates required to perform one cycle of stabiliser check. It can be foreseen that the noise introduced from the extra gates increases the logical error rate of the concatenated code, however, by applying local error detection for each data qubit, we gain extra information concerning the locations of potential errors. This extra information is crucial: we describe an algorithm through which it is translated into modified weights for a minimum-weight matching decoder, allowing superior decisions to be made at the surface code level. Consequently we observe a threshold increase from 1.20\% to 1.42\% with pure phase noise, even assuming that local (pair-wise) and long-range (surface code level) gates have the same fidelity. We then apply the additional, physically-plausible assumption that the computer is structured in such a way that the base-level gates have a high fidelity, and that therefore the local error detection applied to each data qubit has a lower error rate than that in the logical parity check. We observe a further boost of threshold to $2.48\%$ and $4.72\%$ for the error rate ratios of $1:3$ and $1:10$, respectively.

In a real system, phase errors may dominate but other forms of error will also be present at a non-zero level. We model this by having both a pure dephasing process, and a homogeneous depolarising process, simultaneously present with different strengths. Here of course we must note that when there is an asymmetry between $Z$ and $X$ errors, then even with the normal non-concatenated surface code one should change the frequency with which the $X$ and $Z$-stabiliser checks are applied. Therefore for fair comparison, the ratio of $X$ and $Z$-checks is optimised for both codes. We find that our concatenated code yields a higher threshold if the strength of depolarising is smaller than about $70\%$ of the dephasing model's strength, when the long-range gates have the same fidelity as the short-range ones, and can be further increased if the short-range gates have a lower error rate than the long-range ones.

The remainder of this paper is organised as follows. 
In Sec.~\ref{section1}, we introduce the structure of the concatenated surface code. Then, we propose a new surface code decoder in Sec.~\ref{Sec:decoder}, which modifies the minimum weight perfect matching decoder by exploiting the potential error locations observed by the lower-level error detection. Numerical implementation of the concatenated surface code is presented in Sec.~\ref{Sec:numerical}. The paper is concluded in Sec.~\ref{Sec:conclusion} with a discussion of future works.

\section{The concatenated surface code}\label{section1}

\begin{figure}
\begin{centering}
\includegraphics[width=1\columnwidth]{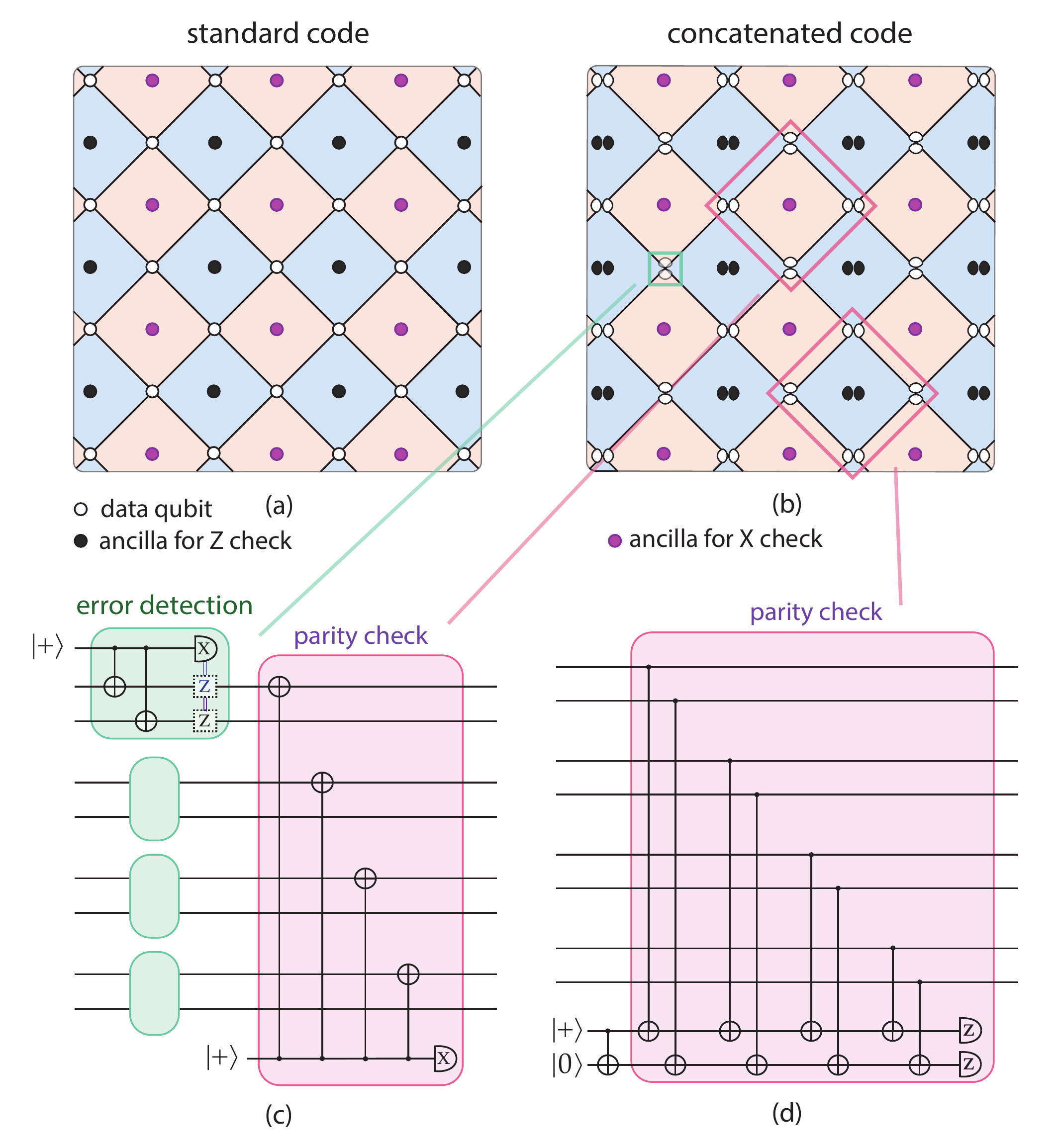}\\
\caption {Schematic view of the surface code and the circuits used for error detection and stabiliser checks. (a) The standard surface code. (b) Our surface code variant. When concatenated with a two-qubit phase detection code, each data qubit becomes a logical qubit that consists two physical qubits (plus one ancilla for error detection, though not shown here). Two physical qubits are required to do a $Z$-stabiliser check, and a single qubit is needed for an $X$-stabiliser check. (c) The circuit for an $X$-check, starting with error detection on each qubit. Once an error is detected, a phase gate is applied to either of the two qubits.(d) The circuit for a $Z$-check, where transversal CNOT gates are applied. In the end, both the two ancillas are measured and the parity of the four qubits is represented by the parity of the measurement outcomes.}
\label{fig:/surfacecode}
\end{centering}
\end{figure}

We concatenate the standard surface code with the two-qubit error detection code, i.e. encode two physical qubits to constitute one data qubit of the surface code, as shown schematically in Fig.~\ref{fig:/surfacecode}. 

\subsection{Standard surface code and two-qubit error detection code}
\label{1.A}
\paragraph{Standard surface code.}
The surface code has a 2D square lattice structure where the data qubits and ancillas sit one next to another. As shown in Fig.~\ref{fig:/surfacecode} (a), in one representation we can locate four data qubits at the edge of each plaquette (the white dots), while the black and purple ones are ancilla qubits located at the center. In this picture, each plaquette defines an operator of either $\bar{X}=X_1X_2X_3X_4$ (with the purple ancilla) or $\bar{Z}=Z_1Z_2Z_3Z_4$ (with the black ancilla), where $X$ and $Z$ are Pauli matrices, referring to the stabiliser generators. When performing the $X$ parity check, a suitable protocol is to perform {\small CNOT} gates from the ancilla qubit to each of the four data qubits, then to measure the ancilla qubit in order to learn the parity of the four data qubits. Consequently, one error on a given data qubit will be identified by two ancilla qubits adjacent to that data qubit. If the error location can be exactly identified, such an error can be corrected by applying another gate of the same type. 

However, stabiliser measurements cannot uniquely determine the error locations and errors in measurements can lead to wrong syndrome outcomes. Therefore, a classical algorithm is used to infer the error locations given the stabiliser measurement information,
 to determine what operations should be performed in order to recover the correct logical state of the quantum system. (Note that in practice, it can suffice to record the corrections that one {\it would} make rather than to actually make then, at least until a non-Clifford operation is scheduled.)  This inference algorithm is referred to as a `decoder'. A number of decoders then have been developed for such a purpose, such as minimum weight perfect matching~\cite{edmonds1965maximum, cook1999computing, fowler2013minimum}, maximum likelihood based on tensor network~\cite{dennis2002topological, bravyi2014efficient, darmawan2017tensor,wang2009threshold, wang2011surface, fowler2012towards,raussendorf2007fault,fowler2009high}, renormalisation group~\cite{bravyi2013quantum, duclos2010fast, watson2015fast} and so on, each with its own advantages and disadvantages. 
 Presently we will describe a decoder we have created for the unique demands of our concatenated code.

\paragraph{Two-qubit error detection code.}
The two-qubit error detection code is the smallest code that can detect one type of single error. We choose the phase detection code because biased phase noise is more commonly seen in experiments, however, for any system with biased $X$ errors, a bit-flip detection code can be adopted under basically the same concept. We employ the encoding $|+\rangle_L=|++\rangle$, and $|-\rangle_L=|--\rangle$. The logical gates are $\bar{X}=XI=IX$, $\bar{Z}=ZZ$, and the stabiliser is $XX$. Any single phase-flip error will be detected when the stabiliser is measured, e.g. by the circuit in the upper left of Fig.~\ref{fig:/surfacecode}(b). However, it will not be possible to determine {\it which} of the two qubits received the phase error; therefore the code is detecting but not correcting. It remains possible to map the defective state back into the code space, by applying a $Z$ operation to either qubit, but this will lead to a logical error $\bar{Z}$ with a significant probability (a $50\%$ probability, if no other information influences our choice -- we discuss this in the Appendix~\ref{parameters}).

\subsection{Concatenated surface code}
As more information concerning error locations is obviosuly beneficial for the decoder to determine the correct recovery operations more accurately, we concatenate the standard surface code with a two-qubit error detection code. Consequently our basic building block, corresponding to the white dots in Fig.~\ref{fig:/surfacecode}(b), is actually three physical qubits: the two encoded qubits and one additional ancilla qubit (not shown in the graph). In the Fig.~\ref{fig:/surfacecode}, the black dots represent the ancilla qubit for a $Z$-stabiliser check of the surface code; it involves two physical qubits. However the ancilla for $X$-checks is different: it is simply a single qubit.

With the concatenation, the number of physical qubits is doubled. Undesirable as it is to increase resource costs, we can expect that the new code may offer advantages because information obtained at the (lower) error detection level can be a powerful resource for acting correctly at the (higher) surface code level. When detecting a phase error, we choose to flip one physical qubit and thus restore the data quit to the proper code space, albeit with the significant probability of having thus implemented an unwanted phase flip on that data qubit. Therefore we record the location of all data qubits where we have observed a phase error; these are now at `high risk' of an error whereas data qubits that have passed the phase-error-detect stage without issue are at `low risk'. This partitioning is very valuable for surface code level inference as we presently discuss. We show the circuit of error detection as the green-shaded area in Fig.~\ref{fig:/surfacecode}(c). 

To do a $Z$-stabiliser check, two ancilla qubits are initialised and prepared to be at $|0\rangle_L$, followed with transversal CNOT gates applied from the data qubit to the ancilla qubits, with the circuit depicted in Fig.~\ref{fig:/surfacecode}(d). Measuring an $X$-stabiliser needs only one ancilla qubit since $\bar{X}=XI=IX$. For simplicity we apply the CNOT gate always on the first qubit, as shown in Fig.~\ref{fig:/surfacecode}(c). The merit of this simplification relies on the data qubit being within its correct two-qubit code space; to maximise this probability we apply the $X$-stabiliser check on all data qubits immediately after the local error detection. Conversely since the $Z$-stabiliser check detects only $X$ errors in the data qubits, we opt not to perform an error detection cycle ahead of it, since this would provide no beneficial information but could add more errors to the system. Thus the overall cycle is: local error-detect, then the $X$-stabiliser checks and finally the $Z$-stabiliser checks, before repeating.

As an aside, we remark that given the circuits in Fig.~\ref{fig:/surfacecode} one might wonder whether the $Z$-check shown on the right can also be performed with four, rather than eight, long range gates, as is done for the $X$-check. Regrettably this would require altering the two-qubit encoding in such a fashion that phase noise from the long range gates cannot be subsequently detected, which would be a net loss to the power of the approach.

\section{Concatenated surface code decoder}
\label{Sec:decoder}

\subsection{Standard surface code decoder}

For the threshold estimates presented in this paper, we use a standard approach of simulating a certain number of full stabilizer cycles, recording classical information at each stage, and finally applying the error inference process, i.e. the decoder, as a single-shot analysis which may either `succeed' or `fail'. (Note that alternatives involving a continuously rolling model also exist in the literature~\cite{fowler2012topological}.) We summarise the canonical approach here, and refer readers to~\cite{nickerson2014freely} as an example of a prior study where the same threshold-finding technique is more fully described.

To estimate a surface code threshold for a given quantum machine, a typical numerical model involves the following stages. Note one does not model the qubits as full quantum entities (which would obviously be exponentially costly in time and memory) but rather one tracks the errors as discrete Pauli events. 
\begin{enumerate}
	\item A total of $n$ imperfect stabiliser checks are modelled, where $n$ is proportional to the size of the code. In each cycle, the full set of surface code stabilisers, $Z$ and $X$ type, are measured. (This can be done simultaneously, if the modelled quantum hardware would permit, or in a set of sub-tasks).
	\item The result of each stabiliser measurement is compared to the previous recorded outcome for that stabiliser -- if it differs then we record that point in time and space as a stabiliser `syndrome event'.
	\item After all cycles are complete, we apply our classical decoder software to analyze the recorded information.
\end{enumerate}
This analysis exploits the observation that any data qubit error, or chain of errors, will lead to two stabiliser syndrome events. 
By successfully sorting all such events into matched pairs, we can infer a proper correction to our surface code state.
 We therefore assign a weight to each potential pairing of events according to their separation (in time and space), with a higher weight indicating that it is less likely that the specific pair is associated with one another.

In order to find the most likely set of pairings, the {\it minimum weight perfect matching} (MWPM) algorithm is used. As the name suggests, the algorithm pairs all events in such a way that the total weight is minimised. Finally, this proposed matching is used to derive a corrective action that should map the entire array back to a correct surface code state; by comparison to the actual record of quantum errors that were introduced in the simulation (which of course would be unknown in a real quantum machine) we are able to record the decoding effort as either a `success' or `failure'. Repeating the entire experiment many times, we determine the percentage success rate. Restarting the complete exercise with a different code size, we discover whether increasing the code size lowers the probability of failure; if so, we are within threshold for quantum error correction.

The minimum weight perfect matching algorithm is commonly used as the algorithm at the heart of decoders for topological codes. Studies based on this approach have  reported high thresholds for the surface code ranging from 0.75\% to 1.4\%, according to the specific variant and error model~\cite{wang2009threshold, wang2011surface, fowler2012towards,raussendorf2007fault,fowler2009high}. In this paper we use it in our decoder for the concatenated code, and we also employ it when we compute results for the standard surface code as a reference.

\subsection{Concatenated surface code decoder}
\paragraph{Wizard decoder.}
The novelty of the present approach is that there is additional information to feed into the classical decoder, in addition to the record of syndrome events. We introduce this by the following thought experiment:
Suppose that we augment the standard surface code with a `wizard' who has the power to detect errors perfectly, and can correct any error with 50\% chance. Whether he corrects it or not, he always records the information into a list; thus half of the list (on average) refers to data qubits with errors, while half refers to those without errors. Note all data qubits that have suffered an error are {\it certainly} on the list. In the remainder of the paper when we refer to `the list' we mean this record of the qubits that are at a high risk of error, here provided by the `wizard' but in practice coming from the error detect circuits. 

During the decoding phase, we have access to this list in addition to our usual syndrome information. We will then only permit pairing of syndrome events that can be connected by a path along which all data qubits lie in the list (we could give infinite `weight' to pairs that cannot be so linked, to prevent our MWPM algorithm from matching them). If the errors are sparsely located, the decoder would then be very powerful -- its pairings are correct with a high probability. In fact, we have confirmed that the threshold data qubit error rate in this circumstance is $59\%$, which is in fact the lattice percolation threshold~\cite{newman2000efficient,jacobsen2014high}: the decoder fails only when the errors are so dense that we can always find a path, connected by listed potential errors, from one boundary to another -- this is a logical error and can not be corrected. Finally note that if, in the above story, the wizard were only able to detect phase errors, then a very high threshold would still be achieved but would apply specifically to errors of that type.

\begin{figure}
\begin{centering}
\includegraphics[width=0.9\columnwidth]{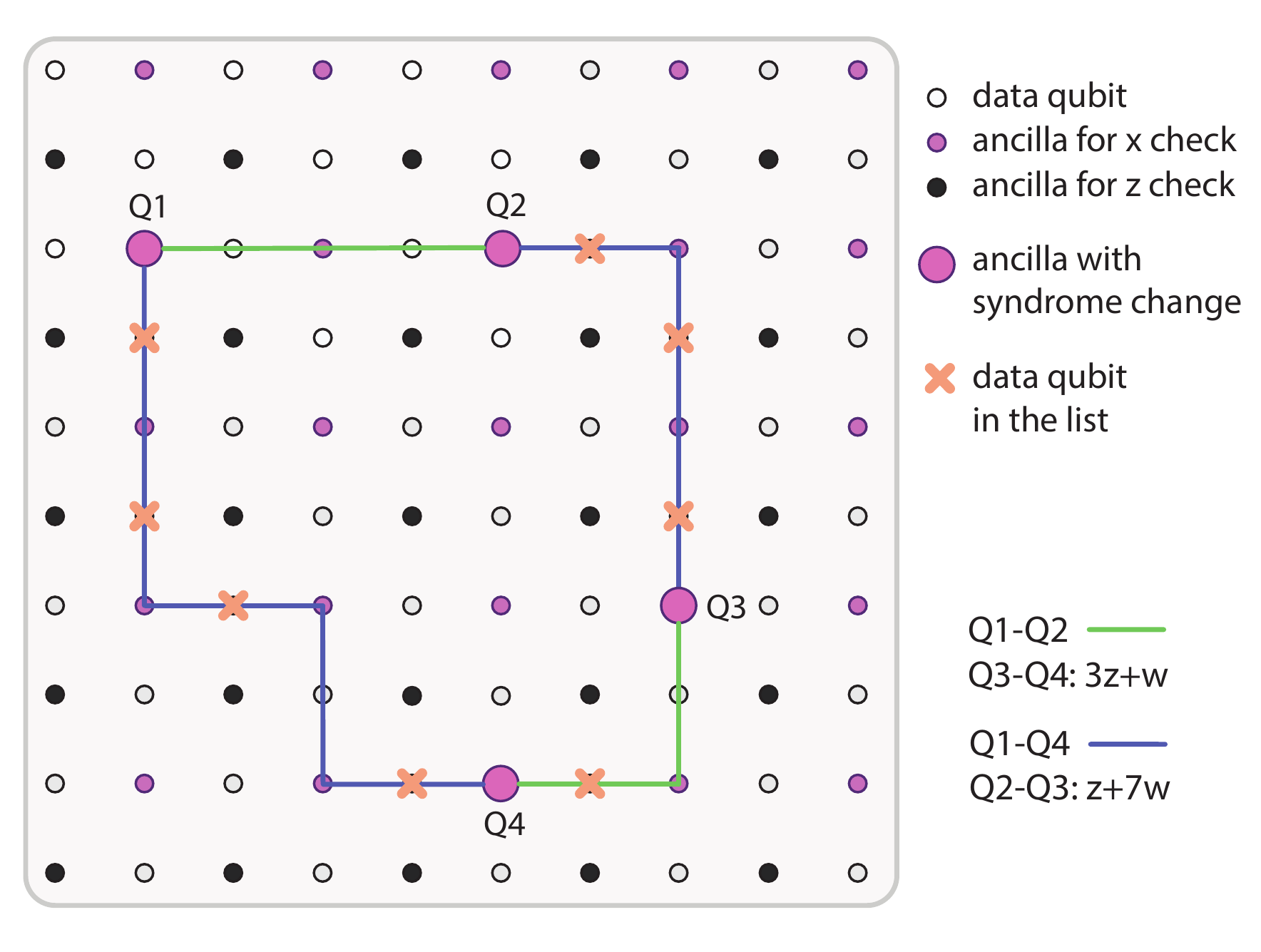}
\caption {Pairing of ancilla qubits based on spatial locations and potentially faulty qubits. All of the qubits shown are in the higher-level surface code. The ancilla qubits with a change of error syndrome are denoted as the big purple dots. The orange crosses represent data qubits identified as likely to have suffered an error according to the lower-level error detection code, i.e. they are on the `list' as described in the main text. Two possible pairings are shown in this graph: Q1\&Q2 and Q3\&Q4, or Q1\&Q4 and Q2\&Q3, connected by the green and purple curves, respectively. The weights for the connections between two adjacent ancilla qubits with and without the listed data qubits are $w$ and $z$. Therefore, while the algorithm based purely on spatial distance will always opt to the first pairing option, the second option gives a lower total weighted distance if $z>3w$.}
\label{fig:/pairing}
\end{centering}
\end{figure}

\paragraph{Realistic decoder.}
For our real concatenated code we do indeed have such a list, which is simply the record of the space/time coordinates where phase errors were detected. However the idealization described above cannot be achieved for two main reasons (1) The error detection process is imperfect due to noise (it can create errors, or incorrectly report the error-status of the pair), (2) It is possible that both physical qubits constituting a data qubit have received errors leading to a logical error which is undetectable. Because of these imperfections we need to permit our decoder to pair syndrome events even when we cannot connect the two through a path along which all
data qubits lie the list. However, we can assign such cases a higher weight, and thus input the knowledge represented by our list in a `softer' form. This is illustrated in Fig.~\ref{fig:/pairing}, which depicts a scenario where we would chose a different pairing than a regular surface code decoder would select because of the additional information from the list. 

{We now describe the approach we have taken to implement a decoder that exploits the valuable, albeit imperfect, information represented by our list. Note that there are many ways to further improve our decoder, but its use provides us with a lower bound on the resulting threshold. That is to say, if we had a large scale quantum computer available now then we could use the decoder exactly as developed for this paper, and then we would expect to realise the performance predicted by our models here; but more likely a superior decoder would be available by the time large scale QIP is possible, in which case we would expect to realise even better performance. }

We apply the Dijkstra's algorithm \cite{dijkstra1959note} which is widely used to efficiently find the shortest paths between nodes in a graph. We employ it to determine a suitable weight for each possible pairing; this replaces the simple `Manhattan distance' calculation that is usually employed in surface code analysis. We can regard each ancilla qubit for $X$-checks as a node, and similarly (but as a separate problem) for $Z$-checks. For each two adjacent nodes, if the data qubit in the middle is on the list, the distance of these two nodes are set to be $w$, otherwise $z$.  
After setting all the distances for each pair of adjacent nodes, Dijkstra's algorithm can determine the minimal distance for each pair of syndrome ancilla qubits and the corresponding path connecting them. It is this `shortest path' that then provides the proper weighting for each pair of syndrome events. Given this weighting, we can use the standard MWMP algorithm to match them. Note that the weight parameters $w$ and $z$ influence the performance of decoder significantly and can be optimized in different practical cases. The ratio $w:z$ is of course dependent on the hardware error rates, with an infinite weight for $z$ being the idealised limit corresponding to the `wizard' in the earlier illustration.  In addition to these spatial weights,  we also consider the syndrome events occurring in different stabiliser cycles, therefore apart from $w$ and $z$, we introduce the time weight $t$. Paths over time are permitted only when the ancilla qubit is connected to itself. The distance of two syndrome events is thus the sum of distance in space and time.

There is a further non-trivial feature related to the use of Dijkstra's algorithm. Since it specifies the entire optimal path connecting each pair, once we have opted to pair two syndrome events then we should correct errors along the lowest weighted path, instead of correcting errors following the shortest spatial path. We have confirmed that doing so has an advantageous effect on the logical error rate. 

We note however that Dijkstra's algorithm is computationally expensive. The complexity is $O(S\cdot n^4/4)$ in our problem where $S$ is the number of syndrome events found in $X$-checks ($Z$-checks). Fortunately, in order to reduce the complexity of Dijkstra's algorithm in a large scale graph, we do not always need to calculate the accurate minimal distance between two syndrome events but instead we can give an approximate distance. Here we predetermine a rather large cut down threshold which is related to the hardware error rates and weight parameters $z,w,t$ and can be optimized in different cases. If the distance between a certain pair of nodes exceed this threshold, the program will stop and  set this value as their distance. The complexity for this modified Dijkstra's algorithm is $O(S\cdot C\cdot n^2/2)$ where $C$ is a constant number related to the cut down threshold, independent of $n$.
This modified Dijkstra's algorithm can improve the efficiency at the expense of the performance of the decoder.

Another important factor in the new decoder is the frequency to apply $X$ and $Z$-checks. As described above, in the case with equal probability of $X$ and $Z$ errors, one should perform equal number of $X$ and $Z$-checks. However, we find that a biased environment in which phase errors dominate necessitates a higher rate of $X$ checks in order to obtain the lowest logical error rate. With the standard surface code, one finds that the ratio of the total number of $X$ and $Z$-checks for the optimal behaviour of the code is roughly the same as the ratio of the probabilities for $Z$ and $X$ errors. Such a discovery is not surprising, as the surface code handles phase and bit errors independently with equal power. Strictly speaking, this is true when the number of basic gate operations involved in an $X$-check is the same as that for a $Z$-check, so that the rate of introduced errors is the same. This will be approximately true in real devices. However, our concatenated code is rather different, not only because it is capable of correcting more phase errors, but also it requires more gates to perform a $Z$-check. In a sense, these two facts have conflicting implications, since when we increase the number of $Z$-checks to achieve balanced performance versus the $X$-checks, more gates are conducted and thus we introduce more errors into the quantum hardware. In our simulation, we find that if the ratio of probabilities for $X$ and $Z$ errors is $\alpha$, then the total number of $X$-checks applied should be somewhat smaller than $1/\alpha$ times of the $Z$-checks applied. We will discuss this further in the Appendix~\ref{parameters}.

\section{Numerical simulation}\label{Sec:numerical}
\subsection{Error models}
In our simulation, we consider a mixture of dephasing and depolarising noise, motivated by their popularity and practical soundness. The noise is stochastic such that each operation can be modelled by a superoperator $\mathcal{N}\mathcal{U}$ with $\mathcal{N}=(1-p)\mathcal{I}+p\mathcal{E}$. Here $\mathcal{U}$ is the ideal operation, and $\mathcal{N}$ is the noisy superoperator with identity channel $\mathcal{I}$ and error channel $\mathcal{E}$ occurring with probabilities $1-p$ and $p$, respectively. For simplicity, we consider the same error rate $p$ for both single- and two-qubit operations.  
For a single qubit, the dephasing error is a $Z$ error and the depolaring error is a uniform mixture of $X$, $Y$, and $Z$ errors. For two qubit gates, the dephasing error is a uniform mixture of errors: $ZI$, $IZ$ and $ZZ$; and the depolaring error is a uniform mixture of the 15 errors: $IX,IY,IZ,XX,XY,\dots,ZZ$. We add noise to every gate, ancilla initialisation and ancilla measurement. 

When mixing the two error models together, we choose the depolarising model and the dephasing model with probabilities $p_{depo}$ and  $p_{deph}$, respectively. That is, the noise superoperator is $\mathcal{N}=(1-p)\mathcal{I}+p(p_{depo}\mathcal{E}_{depo}+p_{deph}\mathcal{E}_{deph})$  with depolarising noise operator $\mathcal{E}_{depo}$ and dephasing noise operator  $\mathcal{E}_{deph}$, and $p_{depo}+p_{deph}=1$. {In the simulation, a gate is applied perfectly with probability $1-p$, otherwise an error is applied with either the depolarising model or the dephasing model based on the biased ratio.} Note this ratio is not the actual ratio of the probabilities of $X$ and $Z$ errors. When the ratio is unity, i.e, when the dephasing model and depolarsing models are applied with an equal chance, the probability for a $Z$ error is roughly $2.8$ times of that for an $X$ error. 

We also consider a quantum hardware with a modular structure, where local gates involved in error detection (green boxes in Fig.~\ref{fig:/surfacecode}) may have a lower error rate than that of the long-range gates involved in the surface code parity checks (red boxes). Therefore we have two overall error rates, $p_d$ and $p_g$, which we refer as the local error rate, and the global error rate, respectively. 
In a summary, the noise superoperator of local and global errors are
\begin{equation}\label{Eq:}
\begin{aligned}
	  \mathcal{N}_d&=(1-p_d)\mathcal{I}+p_d(p_{depo}\mathcal{E}_{depo}+p_{deph}\mathcal{E}_{deph}),\\
	  \mathcal{N}_g&=(1-p_g)\mathcal{I}+p_g(p_{depo}\mathcal{E}_{depo}+p_{deph}\mathcal{E}_{deph}),\\
\end{aligned}
\end{equation}
respectively. In the following, we will consider scenarios with different ratios of global to local error rates $p_g/p_d$, and different ratios of depolarising to dephasing error rates $p_{depo}/p_{deph}$.

\subsection{Simulation results}
We numerically test the capacity of the concatenated code. Each simulation cycle follows the procedure as described before, including stabiliser checks, pairing of ancillas, correction of errors and finally determination of whether there is a logical error. A given experiment is successful if it suffers neither a logical $X$ error nor a logical $Z$, i.e. errors can be perfectly corrected. A Monte-Carlo simulation is applied, with each data point being the average result of at least forty thousand runs. Our main focus here is the threshold of the code under different circumstances, as it is an important measure when comparing two codes, not only because a code with a higher threshold can permit fault tolerant QIP on a more noisy system, but also because the higher-threshold code can be expected to achieve a given target logical error rate with a smaller resource overhead.

\paragraph{Case 1: $p_{deph} = 1$, $p_{depo}=0$, and $p_d\le p_g$. }
To begin with, we consider the case with pure dephasing errors, i.e., $p_{deph} = 1$ and $p_{depo}=0$. Since $X$ errors do not occur in this scenario, we only apply $X$-stabiliser checks for both the concatenated and the standard surface codes for a fair comparison. We first do not distinguish between the fidelity of local and global gates, i.e., with $p_d=p_g$. It is found that the threshold of the concatenated code is 1.42\%, as shown as the first data point in Figure~\ref{fig:/phaseOnly}, higher than that of the standard surface code which is 1.20\% (plots shown in Appendix~\ref{threshold plots}). The result suggests that with only phase noise, the benefits obtained from the extra information exceeds the extra noise introduced from the error detection circuits. 

\begin{figure}[t]
\begin{centering}
\includegraphics[width=1\columnwidth]{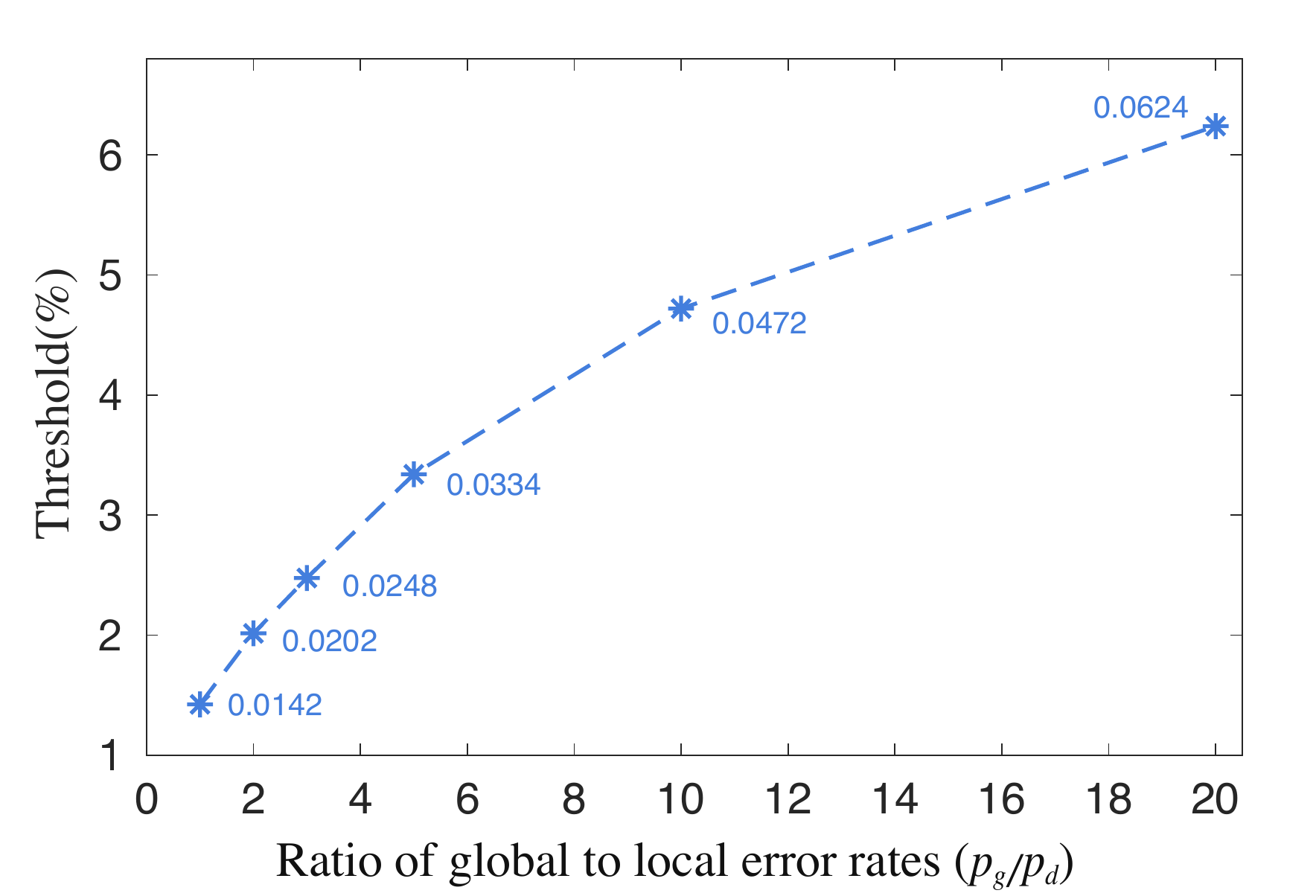}\\
\caption {The increase in noise threshold as we alter the relative error rates for local error detection and global parity check, in a scenario with pure phase errors. The threshold error rate represents the error rate applied for global parity check. A larger ratio leads to an increase of threshold.}
\label{fig:/phaseOnly}
\end{centering}
\end{figure}

We now move to the case that qubits exist in a modular structure such that certain short range two-qubit gates have higher fidelity than other longer range gates, i.e., $p_d\le p_g$. As shown in Fig.~\ref{fig:/phaseOnly}, when we gradually increase the ratio $p_g/p_d$, the threshold error rate is observed to have a continuous gain. Note that the error rate for the determination of threshold is based on $p_g$, as it is larger than $p_d$ and thus plays the more dominant role. We see that the increase of the threshold slows down with a larger ratio $p_g/p_d$. However, if the error detection is perfect, the code should never fail, as no error is introduced to the second qubit (in the circuit in Fig.~\ref{fig:/surfacecode} (c)), while the error detection always identifies the error -- it can only possibly be on the first qubit and can be simply corrected once it is found.

\begin{figure}[t]
\begin{centering}
\includegraphics[width=1\columnwidth]{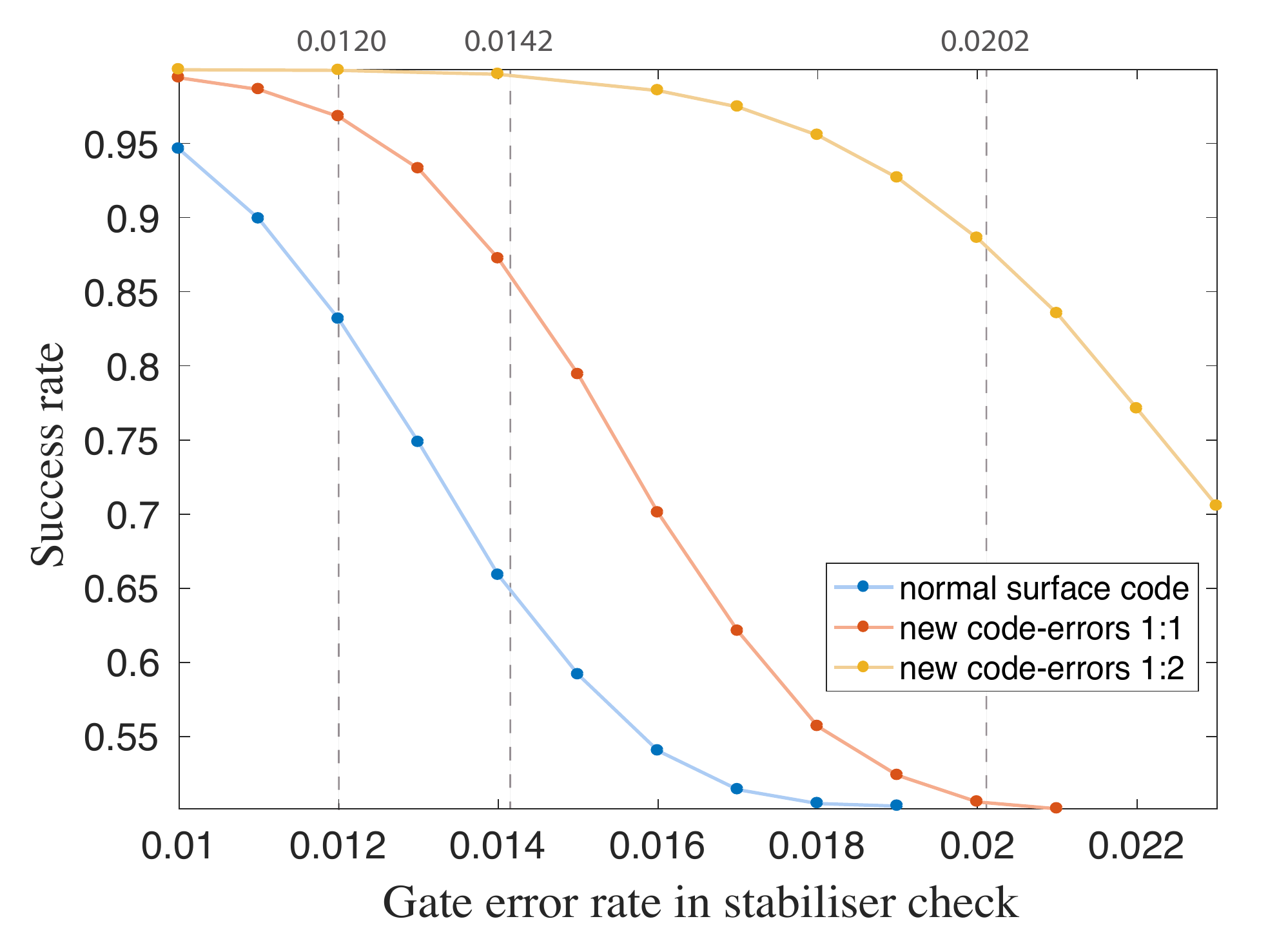}\\
\caption {Code success rate as a change of the gate error rate applied in the parity check cycle under the circumstance of pure phase error. The blue curve represents a standard surface code with a size of 20*20. The orange and yellow curves refer to the concatenated surface code of size 14*14, with $p_d=p_g$ and $p_d=0.5p_g$, respectively. The code sizes are chosen such to make the total number of physical qubits (almost) the same for the two different codes. The grey dashed lines (from left to right) denote where the threshold is, for the blue, orange and yellow curves, respectively. }
\label{fig:/logicalErrorRate}
\end{centering}
\end{figure}

To further assess the performance of the concatenated code versus the standard surface code, we also compare each code's probability of successfully protecting its logical qubit given that each emobdies the same number of physical qubits, e.g, with the same resource requirements. Here we consider the standard surface code with a size of $20*20(=400)$, and the concatenated code of size $14*14$ which actually requires $392$ qubits when we allow for the additional resources needed for our phase error detect layer. Thus the two have nearly the same number of physical qubits. A total of $3*n$ stabiliser cycles are performed, where $n$ is the code size. The result is shown in Figure~\ref{fig:/logicalErrorRate}, where the $y$-axis `success rate' is 1 - logical error rate, and the x axis is the error rate in the parity check cycle. The orange curve represents the concatenated code where error rate for local error detection $p_d$ is the same as the error rate in parity check $p_g$, while the yellow one is the same code but with $p_d = 0.5p_g$. We see that reducing the error rate for error detection yields a large gain in the success rate considering the same error rate of parity check cycle.  Over the whole range, both the curves for the concatenated codes are superior to the blue curve, which corresponds to the standard surface code. The grey dashed curves from left to right indicate the thresholds for the blue, orange and yellow curves as references.

For the concatenated code in the simulations above, as previously discussed, the weights for space and time are adjusted for the highest success rate. We do this empirically though a certain trend has been found and is presented in Appendix~\ref{parameters}. A more rigorous study on the weights is not within the scope of this work but represents a direction for future works.

\paragraph{Case 2:  $p_{depo}/p_{deph}\in[0,1]$. }
So far our comparisons made in the previous simulations are for gates with pure phase noise, now we move on to a more realistic scenario where both phase and bit-flip noise is present but the former still has greater severity. In this case $Z$-stabiliser checks are required to detect $X$ errors, and as mentioned above, in a biased environment, we may employ multiple $X$-stabiliser cycles for each $Z$-stabiliser cycle. For both the standard surface code and the concatenated code, we determined the optimal pattern by trialing different frequencies and selecting the one that leads to the highest success rate. Full details are given in Appendix~\ref{parameters}. Note that for accurate simulations, we need to apply a deep enough simulation (i.e. sufficient number of stabiliser cycles) to ensure that a large number of $X$ errors occur in each run -- otherwise, the results would be skewed by the `edge effect' that the simulation starts from a clean, error-free state.

\begin{figure}
\begin{centering}
\includegraphics[width=1\columnwidth]{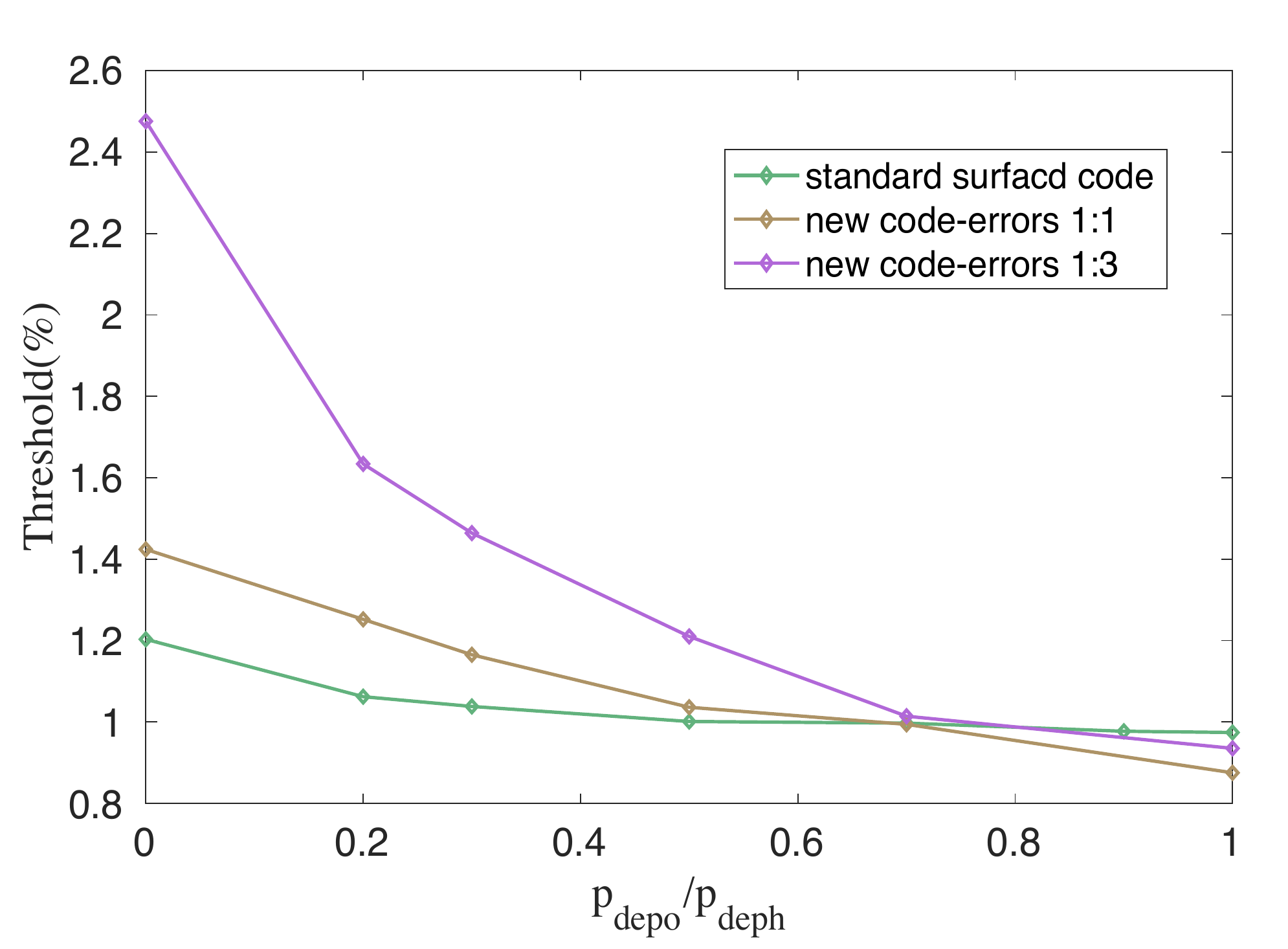}
\caption {The dependence of threshold on the relative strength of dephasing and depolarising errors. The `threshold' quantity along the $y$-axis is the total probability of a physical error; each such error is assigned to one of the two possible models with relative probabilities given by the $x$-axis. The purple (brown) curve corresponds to the concatenated code where the gates in error detection have three times the error rate (the same error rate) as the gates in parity check. For reference the standard surface code is plotted as the green curve.}
\label{fig:/bothError}
\end{centering}
\end{figure}

The data plotted in Fig.~\ref{fig:/bothError} shows the threshold change as we gradually increase the relative strength of the depolarsing versus the dephasing error model. It is not surprising to see that all the three curves representing the threshold, which is the error rate in the stabiliser check cycle, decline as the ratio rises. The green curve referring to the standard surface code starts from the lowest value but goes down slowly. The purple curve stands for the concatenated code whose gate error rate in the error detection is one third of that in the stabiliser check, i.e., $p_d=p_g/3$. We see that it decreases fast and crosses the green curve when the ratio is roughly 0.8, indicating that the concatenated code is no longer beneficial when the strength of depolarsing error is higher than this value. For more interest the concatenated code which has an equal probability for errors happening in error detection and parity check, $p_d=p_g$, is plotted as the brown curve, which as we would expect lies between the other two curves.

\section{Conclusion}\label{Sec:conclusion}
We have proposed a variant of the standard surface code by concatenating it with a simple  two-qubit phase detection code. We developed a new decoder that efficiently takes into account the likely location of errors, obtained from the error detection stage, with a modified minimum weight perfect matching algorithm.
The concatenated code substantially outperforms the canonical surface code (by allowing a significantly higher threshold) for the common scenario where phase error dominates. 
A similar scenario has been considered in Ref.~\cite{tuckett2018ultrahigh}, where a tensor network decoder has been applied to find very high thresholds while assuming perfect active operations. 
Comparison of two works requires a full fault-tolerant analysis of the tensor network decoder considered in Ref.~\cite{tuckett2018ultrahigh}, which is left as an open question.  
It is also an open question to make use of the tensor network decoder or other decoders to efficiently handle the information of the likely location of errors.

The concatenated code becomes yet more advantageous when one considers the likely scenario that local gates (among the groups of three qubits associated with error detection) may have a lower noise rate  compared to the gates that link between such groups. 
Such a scenario would correspond to modular hardware with exquisite local quantum control and more noisy global links. Modular quantum computing  has been extensively studied  and shown to be feasible even with near-term hardware  \cite{nickerson2014freely,li2016hierarchical}. Our work for the first time studied the concatenation of the surface code with a lower-level error-detect code, in the context of modular hardware. It remains an interesting question to instead consider other codes for local modules.  For instance, one can consider the four-qubit error detection code, which can detect both single phase and bit errors, and combine it with a decoding algorithm similar to the one described here. 

\section{Acknowledgements}
SCB and XX are supported by the Office of the Director of National Intelligence (ODNI), Intelligence Advanced Research Projects Activity (IARPA), via the U.S. Army Research Office Grant No. W911NF-16-1-0070. The views and conclusions contained herein are those of the authors and should not be interpreted as necessarily representing the official policies or endorsements, either expressed or implied, of the ODNI, IARPA, or the U.S. Government. The U.S. Government is authorized to reproduce and distribute reprints for Governmental purposes notwithstanding any copyright annotation thereon. Any opinions, findings, and conclusions or recommendations expressed in this material are those of the author(s) and do not necessarily reflect the view of the U.S. Army Research Office. QZ acknowledges support by the National Natural Science Foundation of China Grant No. 11674193.
SCB and XY acknowledge EPSRC grant EP/M013243/1. The authors would like to acknowledge the use of the University of Oxford Advanced Research Computing (ARC) facility in carrying out this work. http://dx.doi.org/10.5281/zenodo.22558
\bibliography{surfacecode}
\appendix

\section{Parameters considered in the decoder}
\label{parameters}
A couple of parameters have been considered regarding different circumstances.
\subsection{The correction to apply on the physical qubit}
When the error detection finds a single error, a phase gate will be applied to a physical qubit. As discussed in the main text, the extra phase gate will either cancel out the error or create a logical error to the data qubit. If no other information is provided, one may apply the phase gate on one of the two physical qubits with equal probability. However, a trend can be found based on the specific structure of the circuit used for surface code parity checks (see Fig.~\ref{fig:/surfacecode}): since the gate for parity checks is always applied on the first physical qubit, noise on the second physical qubit can only be introduced from the error detection gates. Intuitively, if an error is detected, we could estimate where it is depending on the relative error strength $p_d/p_g$. If the low-level gates have a much higher fidelity, the error generated in the last parity check cycle (on the first qubit) is likely to be identified correctly, thus under such circumstance the phase gate should be applied to the first qubit. On the other hand, if the low-level gates have a comparable error rate as the long-range gates, we have the following estimate. With the dephasing error model, the probability for a phase error occurring in either qubits is $\frac{2}{3}p_d$, and for measurement error is $\frac{10}{3}p_d$, indicating reasonable chance of a wrong error syndrome, with which, it will be safer to apply the phase gate to the second qubit, as such it will not be detected by the parity check and can be corrected in the next cycle. In the simulation, it is found that when ${p_d}/{p_g}$ is smaller than 0.5, one should apply the phase gate to the first qubit, while if not, the gate should be placed in the second qubit.

\subsection{The weights for time and space}
As introduced in the main text, the distance between any two syndrome events is the sum of their distance over time and space. In the calculation, the three weights $w,z,t$ are based on the extra information obtained from error detection.

For convenience we make the weight $w$=1, and refer to $t$ and $z$ as time weight and space weight, respectively. Table~\ref{table1} shows the weights used in the simulations. The first two columns correspond to the case where the phase gate is applied on the second physical qubit: the distinct difference of weights compared with the others is elusive, however, we found that the space weight is not independent as we introduce the cut down threshold as another variable into the decoder. In the calculation of the distance of the two syndrome events, if the distance found is higher than this threshold, the program will stop and make the distance infinity, as to not pair the two syndrome events. By doing so we shorten the time cost for each simulation run, and we found it also constrains the space weight --- which should be smaller than the threshold so as to permit the connection of two data qubits that are not in the list. 

\begin{table}[t]
\begin{tabular}{ccccccc}
\hline\hline
$p_d/p_g$      & 1  & 1/2 & 1/3  & 1/5 & 1/10 & 1/20 \\ \hline
\textbf{Space weight $z$} & 12  & 12  & 4.5  & 4   & 3    & 3    \\ 
\textbf{Time weight $t$}  & 3.5 & 3.5 & 0.85 & 0.5 & 0.4  & 0.35 \\ 
\textbf{Cut down} & \multirow{2}{*}{$3n+1$}& \multirow{2}{*}{$3n+1$}& \multirow{2}{*}{$2n+4$} & \multirow{2}{*}{$n+10$} & \multirow{2}{*}{$n+6$}  & \multirow{2}{*}{$n+5$} \\
\textbf{threshold}  &  & &&&& \\ \hline\hline
\end{tabular}
\caption {Different weights and cut down threshold used in the simulations. $n$ represents the size of the code.}
\label{table1}
\end{table}

\subsection{The approximate Dijkstra Algorithm}

We denote all the nodes in the graph as $G$ and the single source start point as $s_0$, the vector $dist[s]$ as the distance between $s$ and $s_0$, $prev[s]$ as the previous node adjacent to $s$ in the path connecting $s$ and $s_0$ with distance $dist[s]$. $S$ and $Q$ denote the sets containing the nodes visited and unvisited, respectively. $c$ denotes the predetermined cut down threshold. The detailed procedure for single source Dijkstra Algorithm is shown in Algorithm 1. This procedure can only give the distances starting from one single node. Thus in our case we repeat this algorithm a few times starting from different syndrome change nodes. 

\begin{algorithm}[H]
\begin{algorithmic}[1]
\State For any node $s\in G/\{s_0\}$, $dist[s] =\infty$, $prev[s]=s_0$, $dist[s_0] = 0$
\State{ $S=\{s_0\}$, $Q=G\{s_0\}$ }
	\While {$Q\neq \varnothing$}
	\State {u=Extract-Min(Q), S.insert(u)}
	\If {$dist[u]<c$}
	\For {any $v\in Q$ adjacent to $u$}
        \If {$dist[v]>dist[u]+weight[u,v]$}
       	\State {$dist[v]=dist[u]+weight[u,v]$, $prev[v]=u$}
       	\EndIf
	\EndFor
	\EndIf
    \EndWhile
\end{algorithmic}
	\caption{Approximate Dijkstra Algorithm}
\end{algorithm}

\subsection{The frequency to apply $X$ and $Z$ checks}
When mixing the depolarising model with the dephasing model, the pattern for applying $X$ and $Z$-stabiliser checks requires adjustment. The reason to do so stems from the fact that the overall success rate is the product of the success rates for logical $Z$ and $X$ errors, thus the highest success rate is achieved when the success rate for the two types of logical errors is the same. Therefore, with a biased error model, more $X$-checks are to be performed than $Z$-checks.

The frequencies of the $X/Z$-stabiliser checks considered in our simulation are shown in  Table~\ref{table2}, which generates the lowest overall logical error rate. Here, the frequency $F$ refers to the rounds of $X$-stabiliser checks before one round of $Z$-stabiliser check is applied. Unsurprisingly as the relative strength of depolarising error reduces, the frequency increases. With the standard surface code, the optimal frequency is roughly the same as the rounded relative probability to find a phase and bit-flip error, as indicated in the last row of the table. Note the frequency may not be strictly ideal, since we do not consider the case when it is a non-integer (e.g. 3/2 means three rounds of $X$-checks followed with two rounds of $Z$-checks), as it may change the error pattern significantly. As for the concatenated code, smaller frequencies are observed as the $p_d/p_g$ increases. It can be explained by the fact that as $p_d$ becomes larger, more errors are introduced during the error detection cycle which increases the logical error rate. Given that the total number of stabiliser checks is fixed, a smaller number of $X$-checks is therefore more beneficial. 

\begin{table}[H]
\begin{tabular}{c|ccccccc}
\hline\hline
\multicolumn{2}{l}{Relative strength} & 0.2 & 0.3 & 0.5 & 0.7 & 0.9 & 1.0 \\ \hline
\multirow{3}{*}{F} & standard code& 9  & 6   & 4  & 4   & 3   & 3    \\ 
                   & $p_d=1/3p_g$ & 5   & 3   & 2   & 2   & 2   & 1    \\ 
                   & $p_d=p_g$   & 4   & 3  & 2   & 1   & 1   & 1    \\ \hline
\multicolumn{2}{l}{$Z/X$ error rate}  & 9.918 & 6.738 & 4.352 & 3.597 & 2.958 & 2.786 \\ \hline\hline             
\end{tabular}
\caption{The optimal frequencies of $X/Z$ checks for the concatenated and the standard codes as the relative strength of depolarising and dephasing noise changes. F refers to frequency, e.g. how many rounds of $X$-stabiliser checks are applied before one round of $Z$-stabiliser checks. The $Z/X$ error rate is the relative probability for $Z$ and $X$ errors happening in the standard surface code, when the frequencies of $X$ and $Z$ checks are as above.}
\label{table2}
\end{table}

\section{Threshold plots for the case with pure phase errors}
\label{threshold plots}

\begin{figure*}[htbp]
\begin{centering}
\includegraphics[width=2\columnwidth]{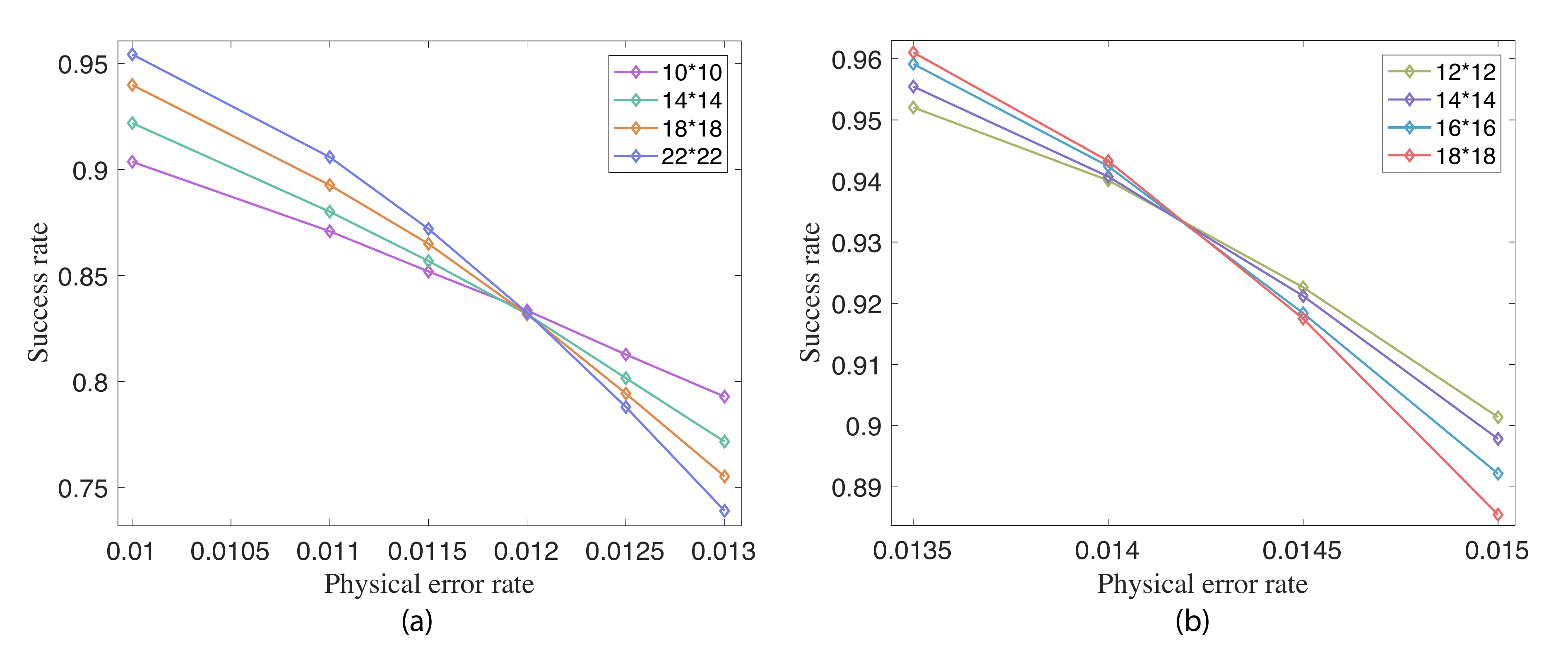}
\caption {Dependence of the logical success rate with the physical error rate and the size of the codes for (a) the standard surface code and (b) the concatenated code. The error rate in the error detection cycle is the same as in the parity check cycle. The crossing point of the curves defines the threshold of that code. We see that when only phase error is present, the standard surface code has a threshold of around 1.20\%, smaller than that of the concatenated code, which is around 1.42\%.}
\label{fig:/plot}
\end{centering}
\end{figure*}

Here we show the threshold plots with the standard surface code and the concatenated code for comparison. Only phase error is present in the simulation, and we make the pessimistic assumption that the local gates for error detection have the same fidelity as the long-range gates for parity check. As shown in Fig.~\ref{fig:/plot}, the threshold of the standard surface code is about 1.20\%, appreciably smaller than the threshold of the concatenated code (1.42\%).

\end{document}